\documentclass[pss]{wiley2sp}

\usepackage[english]{babel}%
\usepackage{bm}%
\usepackage{graphicx}
\usepackage[active]{srcltx}
\usepackage{amsmath}
\usepackage{amssymb}

\newcommand{\mean}[1]{\mathord{\left\langle #1 \right\rangle}}

\newcommand{\llangle}{\mathord{\langle\!\langle}}
\newcommand{\rrangle}{\mathord{\rangle\!\rangle}}

\newcommand{\halb}{\mathord{\frac{1}{2}}}

\newcommand{\fk}{\mathord{\mathbf k}}

\newcommand{\fR}{\mathord{\mathbf R}}
\newcommand{\fS}{\mathord{\mathbf S}}
\newcommand{\fs}{\mathord{\boldsymbol \sigma}}
\newcommand{\ek}{\mathord{\epsilon(\mathbf k)}}

\newcommand{\intmax}{\mathord{\int\limits^{+\infty}_{-\infty}}}
\newcommand{\zs}{\mathord{z_{\sigma}}}
\newcommand{\comm}[1]{\mathord{\left[ #1 \right]_-}}

\newcommand{\dt}{\mathord{\frac{\partial}{\partial T}}}

\begin{document}

\title{Curie temperatures of the concentrated and diluted Kondo-lattice model as a possible candidate to describe magnetic semiconductors and metals} 
\titlerunning{Concentrated and diluted Kondo-Lattice model}

\author{M. Stier\textsuperscript{\textsf{\bfseries \Ast ,1}} and W. Nolting\textsuperscript{\textsf{\bfseries 1}}}
\mail{\textsf{stier@physik.hu-berlin.de}}
\authorrunning{M. Stier et al.}

\institute{\textsuperscript{1}Festk\"orpertheorie, Institut f\"ur Physik, Humboldt-Universit\"at, 12489 Berlin, Germany}
\received{}
\published{}


\abstract{
\abstcol{%

We present a theory to model carrier mediated ferromagnetism in concentrated or diluted local moment systems. The electronic subsystem of the Kondo lattice model is described by a combined equation of motion / coherent potential approximation method. Doing this we can calculate the free energy of the system and its minimum according to the magnetization of the local moments. Thus also the Curie temperature can be determined and its dependence on important model parameters. We get qualitative agreement with the Curie temperatures' experimental values of Ga$_{1-x}$Mn$_x$As for a proper hole density.
}
{}
}

\maketitle

\section{Introduction}

Ferromagnetism is a solid state phenomenon which arises from the interplay of magnetically active atoms or their electrons, respectively. Naively one would think that high Curie temperatures are connected to a high density of those atoms. Thus it was suprising when relatively high Curie temperatures\cite{ohno1,ohno2,wang} in diluted magnetic semiconductors (DMS) with $T_Cs$ over 100K were discovered - even at small amounts of magnetic atoms. This finding and the high spin polarization of the carriers made the DMS a possible candidate for spintronic devices\cite{spintr}. It is a major task to understand these materials and thereby improve them to eventually achieve room temperature magnetism.\\
One prominent DMS is Ga$_{1-x}$Mn$_x$As which has a $T_C$ up to $\approx173$K for a concentration of $x=9\%$ of manganese ions\cite{dmsMF}. Manganese, which replaces dominantly gallium, has a partly filled $3d^5$-shell, in contrast to gallium's completely filled $3d^{10}$-shell. To calculate the effect of this doping several methods can be used. Ab-initio calculations are able to calculate the full density states of a specific material. This gives insight into the position of the impurity levels within the bulk material. Additionally Curie temperatures of different DMS can be calculated accurately\cite{sato,sato2}. Besides ab-initio methods, model studies are useful to find the main mechanism which leads to ferromagnetism in DMS. For Ga$_{1-x}$Mn$_x$As the $3d$-states of manganese lie deep below the valence band\cite{spintr,sato} of the host material and are therefore assumed to lead to a localized spins $\fS_i$ at the impurity site $\fR_i$ with spin quantum number $S=5/2$. Although the states are deep below the Fermi level they can interact with carriers (holes) in the valence band which influences its band structure. This results in an effective exchange between the localized spins $\fS_i$ (even though they are far away from each other at strong dilutions) which can lead to ferromagnetic order.\\
One of the major problems is to have control over the carrier density ($n$ for electrons or $p$ for holes). Nominally one manganese ion provides one hole but due to compensation effects the ratio $\gamma=p/x$ is usually smaller than one. As the mean-field Zener model predicts $T_C\sim p^{\frac{1}{3}}$  this seems to reduce the Curie temperature. In contradiction experiments show\cite{expcomp,expcomp2} that a hole compensation is rather necessary for a finite $T_C$. The same is true for theoretical approaches beyond the mean-field approximation\cite{tang,bouz2007,popesc}. To calculate $T_C$ the Kondo lattice model (KLM) is often mapped onto a Heisenberg model, which is then solved with different approximation methods. Contrary to that we propose a procedure how to calculate the free energy exclusively from the electronic subsystem so that we get the Curie temperature directly from the KLM.\\
The paper is organized as follows. First we introduce the model Hamiltonian. After that we describe how we calculate the internal energy of the system by an equation of motion approach. In the case of a diluted system we combine that with a coherent potential approximation. As the next step we use the internal energy to get the free energy where we also need the entropy at $T=0$K. From the free energy we calculate Curie temperatures in dependence of several model parameters. These are compared with experimental data.

\section{Model}

We use the Kondo lattice model to describe the interactions between local moments $\fS_i$ and itinerant conduction electrons with spin $\fs_i$, with the standard Hamiltonian
\begin{align}
\mathcal H =&H_0+H_{sf}\nonumber\\
=&\sum_{\mean{i,j}\sigma}T_{ij}c^+_{i\sigma}c_{j\sigma}-J\sum_{i} \fS_i\cdot\fs_i\ .\label{eqho1}
\end{align}
The itinerant electrons, represented by the creators (annihilators) $c_{i\sigma}^{(+)}$, move from lattice site $\fR_i$ to $\fR_j$ due to the hopping $T_{ij}$. They prefer to align parallel ($J>0$) or antiparallel ($J<0$) to the local moment on the same site.\\
In the context of DMS the localized moments are formed by the $d$-states of the magnetic impurities while the itinerant carriers move in the valence band (holes) or in the conduction band (electrons). In Ga$_{1-x}$Mn$_x$As the hole created by the replacement of Ga with Mn is within the three antibonding $sp-d$-levels with dominant As 4p character\cite{spintr}. This actually requires a multiband treatment but to simplify the calculations the valence band is approximated in a one-band model. The conduction band is completely neglected.\\
Since the magnetic impurities are randomly distributed within the host material we have to include disorder to our model. This is done via an introduction of  a random variable $x_i=0,1$ indicating if a magnetic impurity is present at lattice site $\fR_i$. The concentration $x$ of the magnetic atoms is fixed by the requirement $x=\frac{1}{N}\sum_i x_i$. Besides the magnetism of the impurities there can be other effects on the electrons\cite{popescu,bouz2007}, e.g. different atomic levels $T_0^{\alpha}$ for magnetic ($\alpha=$M) and non-magnetic sites ($\alpha=$NM). We choose $T_0^{\text{M}}=T_{ii}^{\text{M}}=0$ while  $T_0^{\text{NM}}=T_{ii}^{\text{NM}}$ may be unequal zero. We end up with the Hamiltonian
\begin{align}
 \mathcal H =& \sum_{\mean{i,j}\sigma}T_{ij}c^+_{i\sigma}c_{j\sigma}+\sum_i(1-x_i) T_0^{\text{NM}}n_{i\sigma}-\nonumber\\
&-\frac{J}{2}\sum_{i\sigma}x_i\left(\zs S_i^zn_{i\sigma} +S_i^{\bar\sigma}c^+_{i\sigma}c_{i\bar\sigma}\right)\label{eqH2}
\end{align}
where we wrote the scalar product in (\ref{eqho1}) with the help of Pauli matrices $\hat\fs$ and $\fs_i =\halb\sum_{\sigma\sigma'}c^+_{i\sigma}\hat\fs_{\sigma\sigma'}c_{i\sigma'}$ explicitly. There is now an Ising term $\zs S_i^zn_{i\sigma}$ ($\zs=\pm1$, $n_{i\sigma}=c^+_{i\sigma}c_{i\sigma}$) and a spinflip part $S^{\bar\sigma}_ic^+_{i\sigma}c_{i\bar\sigma}$ ($S_i^{\sigma}=S^x_i+\zs iS^y_i$, $\bar\sigma=-\sigma$).\\
Since our model is not restricted to small doping $x$ we are able to calculate the whole range $x=0\dots 1$ to show general trends of the influence of dilution.

\section{\label{secGF}Electronic part}

In this work the key quantity is the internal energy $U=\mean{\mathcal H}$. It can be derived in the KLM from the single-electron Green's function (SE-GF) 
\begin{align}
G_{ij\sigma}(t-t')=& \llangle c_{i\sigma}(t), c^+_{j\sigma}(t')\rrangle\nonumber\\ =&-i\Theta(t-t')\mean{[c_{i\sigma}(t),c^+_{j\sigma}(t')]_{\pm}}
\end{align}
with the thermodynamic average $\mean{\dots}$ and the usual step function $\Theta(t-t')$. Operators are in the time dependent Heisenberg representation. The GF can be Fourier-transformed to its energy representation
\begin{align}
G_{ij\sigma}(E)=&\int d(t-t')G_{ij\sigma}(t-t')e^{iE(t-t')}\\
 =&\llangle c_{i\sigma};c^+_{j\sigma}\rrangle\ .\nonumber
\end{align}
In the KLM the internal energy then is simply given as
\begin{align}
 \frac{U}{N}=-\frac{1}{N\pi}\sum_{i\sigma}\intmax dE\ E f_-(E)\text{Im}G_{ii\sigma}(E)\ ,\label{eqU}
\end{align}
where the Fermi function $f_-(E)=(e^{\beta(E-\mu)}+1)^{-1}$ contains the chemical potential $\mu$ and the inverse temperature $\beta^{-1}=k_B T$. Thus, to calculate $U$, we need to find an expression for $G_{ij\sigma}(E)$. In a first step this will be done for the concentrated lattice ($x_i=1, \forall i$) and then will be extended to diluted lattices.

\subsection{\label{secGFc}Single electron Green's function in the concentrated lattice}

To get the SE-GF we will perform a moment conserving decoupling approach (MCDA\cite{mcda1997}) of higher Green's functions appearing in several equations of motion (EOM). Let us start with the EOM of $G_{ij\sigma}$, which reads
\begin{align}
 \sum_l\left(E\delta_{il}-T_{il}\right)G_{lj\sigma}(E)=\delta_{ij}+\llangle [c_{i\sigma},H_{sf}]_-;c^+_{j\sigma}\rrangle\ .\label{eomGF1}
\end{align}
We can introduce a self-energy
\begin{align}
 \llangle [c_{i\sigma},H_{sf}]_-;c^+_{j\sigma}\rrangle\equiv \sum_l M_{il\sigma}(E)G_{lj\sigma}(E)\label{eqself}
\end{align}
which leads, after Fourier transformation, formally to the solution
\begin{align}
 G_{\fk\sigma}(E)=\frac{1}{E-\ek-M_{\fk\sigma}(E)}\ .\label{eqGF}
\end{align}
The commutator in (\ref{eqself}) can also be calculated explicitly, yielding
\begin{align}
 \llangle [c_{i\sigma},H_{sf}]_-;c^+_{j\sigma}\rrangle=&-\frac{J}{2}\left(\zs\mean{S^z}G_{ij\sigma}(E)\right.+\label{eqFG}\\ &\left.+\zs\Gamma_{iij\sigma}(E)+F_{iij\sigma}(E)\right)\nonumber\\
\Gamma_{ikj\sigma}(E)=&\llangle \underbrace{(S^z_i-\mean{S^z_i})}_{\delta S_i^z}c_{k\sigma};c^+_{j\sigma}\rrangle\\
F_{ikj\sigma}(E)=&\llangle S^{\bar\sigma}_ic_{k\bar\sigma};c^+_{j\sigma}\rrangle
\end{align}
with the reduced Ising function $\Gamma_{ikj\sigma}(E)$ and the spinflip (SF) function $F_{ikj\sigma}(E)$. These functions can also be calculated by their EOMs. Within these equations we treat diagonal ($i=k$) and non-diagonal terms ($i\neq k$) differently. For the non-diagonal elements we replace approximately
 \begin{equation}
 \comm{c_{i\pm \sigma}, H_{sf}} \rightarrow \sum_l M_{il\pm\sigma}(E)c_{l\pm\sigma}
\end{equation}
as in Eq. (\ref{eqself}), which is only strict for $G_{ij\pm\sigma}(E)$, of course. As an example:
\begin{align}
 \llangle S_i^z\comm{c_{k\pm \sigma}, H_{sf}};c_{j\sigma}^+\rrangle\approx\sum_l M_{kl\sigma}\llangle S_i^zc_{l\sigma};c_{j\sigma}^+\rrangle\ .
\end{align}
The higher Green's function is therewith traced back to a linear combination of Ising functions. It is easy to see that the functions on the left- and the right-hand side have the same pole structure differing only by the respective spectral weights. The latter is approximately compensated by a self-consistent determination of the expansion coefficients which are just the elements of the central self-energy matrix. Higher Green's functions with commutators containing only spin operators (e.g. $\comm{S_i^z,H_{sf}}$), are neglected, since they scale with magnon energies which are much lower than those of the electrons\cite{mcda1997}. The commutators of the diagonal terms, however, are calculated explicitly which leads to further Green's functions $\llangle A_{i\sigma}c_{i\pm\sigma};c^+_{j\sigma}\rrangle$, where $A_{i\sigma}$ stands for different combinations of spin and electron operators. These functions can be expressed in some limiting cases (e.g. $S=\halb$, $\mean{S^z}=S$, $n=0$ or $n=2$) rigorously by the three Green's functions mentioned above suggesting the general ansatz
\begin{align}
 \llangle A_{i\sigma}c_{i\pm\sigma};c^+_{j\sigma}\rrangle =&a_{A\sigma}G_{ij\sigma}(E)+\nonumber\\
&+b_{A\sigma}\Gamma_{iij\sigma}(E)+c_{A\sigma}F_{iij\sigma}(E)
\end{align}
which we approximate to be valid for all parameter constellations. The prefactors $a_{A\sigma},\dots$ are fixed by the requirement of conservation of the first spectral moments of the Green's functions and their according expectation values. Since the moments are conserved we then get a correct high energy behavior of these functions\cite{ISA1,modpert}.\\
After performing the mentioned approximations we get unique solutions for the Ising- and spinflip Green's functions (details in Ref.\cite{mcda1997}). These can be written in the form
 \begin{equation}
 X_{iij\sigma}(E)=C_{X\sigma}(E) G_{ij\sigma}(E)\qquad, X=\Gamma,F\label{eqgensol}
\end{equation}
and we see that they are proportional to the one-electron Green's function. This allows us, by comparing Eqs. (\ref{eomGF1}, \ref{eqself}) and (\ref{eqFG}), to identify the prefactors $C_{X\sigma}(E)$ in (\ref{eqgensol}) with the self-energy
\begin{align}
 M_{\sigma}(E)=&\underbrace{-\frac{J}{2}\zs\mean{S^z}}_{M_{\sigma}^{\text{mean-field}}}\underbrace{-\frac{J}{2}\left(\zs C_{\Gamma\sigma}(E)+C_{F\sigma}(E)\right)}_{M_{\sigma}^{\text{many-body}}(E)}\ ,\label{eqself2}
\end{align}
which splits into a mean-field and a more complicated "many-body" part. The latter is defined by the $C_{X\sigma}(E)$, which depend on several expection values. These either contain only spin operators or also electronic operators. The "pure spin operator" values (e.g. $\mean{(S^z)^2}$) can all be expressed by the well known formula of Callen\cite{callen} as functions of the local moment magnetization $\mean{S^z}$. To do this we implicitly assume that a mapping of the KLM onto a Heisenberg model is in principle possible, but we do not have to perform it explicitly. All other expectation values (e.g. $\mean{n_{i\sigma}}$, $\mean{S^{\sigma}_i c^+_{i-\sigma}c_{i\sigma}}$, $\mean{S^z_in_{i\sigma}}$) can be calculated with the spectral theorem from the known Green's functions $G_{ij\sigma}$, $\Gamma_{iij\sigma}$ and $F_{iij\sigma}$. But the $C_{X\sigma}(E)$ do not only depend on scalars but also on the local Green's function $G_{\pm\sigma}(E)=1/N\sum_{\fk}G_{\fk\pm\sigma}(E)$ and the self-energy $M_{\pm\sigma}(E)$ itself. These dependencies necessitate a self-consistent calculation of the self-energy.\\
Especially the dependence on the local Green's function
\begin{align}
 M_{\sigma}(E)=M_{\sigma}(E,G_{\pm\sigma}(E))\label{eqgdepnd}
\end{align}
will play a crucial role when we consider diluted systems.\\
Since we want to focus on the interaction of local moments and carriers, we have neglected a direct influence of the Coulomb repulsion in the conduction band. An addition of that would be especially necessary to describe narrow band materials like the manganites\cite{heldvoll}. The inclusion of a Hubbard-type Coulomb term could lead to extra effects in magnetic ordering (band-magnetism), which are not accounted for in this work.

\subsection{\label{secGFd}Single electron Green's function in the diluted lattice}

The diluted KLM can be described as a binary alloy. The components are defined by the sites with or without magnetic impurities. In the coherent potential approximation (CPA\cite{velicky}) the configurationally averaged Green's function of the \emph{total} system is given by 
\begin{equation}
 \mean{G_{\fk\sigma}(E)}=\left(E-\ek-\Sigma_{\sigma}(E)\right)^{-1}\ .\label{eqgcpa}
\end{equation}
The CPA self-energy $\Sigma_{\sigma}(E)$ can be derived self-consistently from
\begin{align}
 0 = &x	 \frac{\eta^{\mathrm{M}}_{\sigma}-\Sigma^{}_{\sigma}}
		      {1-(\eta^{\mathrm{M}}_{\sigma}-\Sigma^{}_{\sigma})\mean{G^{}_{ii\sigma}}}+\nonumber\\
  &+ (1-x)\frac{\eta^{\mathrm{NM}}_{\sigma}-\Sigma^{}_{\sigma}}
		      {1-(\eta^{\mathrm{NM}}_{\sigma}-\Sigma^{}_{\sigma})\mean{G^{}_{ii\sigma}}}.
\label{eq:cpa_selfconsistent}
\end{align}
As the concentration of the magnetic impurities $x$ is a parameter we only need to determine the potentials $\eta^{\mathrm{M},\mathrm{NM}}_{\sigma}$ to solve the problem. We set $\eta^{\mathrm{NM}}_{\sigma}=T_0^{\text{NM}}$ for the non-magnetic sites. For the magnetic sites we choose the self-energy (\ref{eqself2}) as an energy dependent potential $\eta^{\text{M}}_{\sigma}(E)=M_{\sigma}(E)$ (dynamical alloy analogy, DAA\cite{takahashi,tang}).  In Section \ref{secGFc} we calculated the self-energy $M_{\sigma}(E)$, put it into formula (\ref{eqGF}) to get $G_{\fk\sigma}(E)$ and calculate $M_{\sigma}(E)$ again until self-consistency is achieved. But we have to keep in mind that the self-energy was derived for a \emph{concentrated} lattice. In a diluted system the magnetic component is embedded in the total system. Thus we change the self-consistency cycle a bit.\\
Since the main change of the Hamiltonian (\ref{eqH2}) for the diluted system is only the inclusion of scalar random variables $x_i$, we assume that the equations of motion of the magnetic subsystem have the same formal structure as in Section \ref{secGFc}. That means especially that the self-energy $M_{\sigma}(E)$ is a function of the local Green's function $G_{\pm\sigma}(E)$ (Eq. (\ref{eqgdepnd})). As we want to calculate only the self-energy of the magnetic subsystem it should only depend on the local Green's function of this subsystem. Thus we replace $G_{\pm\sigma}(E)\rightarrow G^{(\text{M})}_{\pm\sigma}(E)$ in Eq. (\ref{eqgdepnd}) yielding
\begin{align}
 M_{\sigma}(E,G^{(\text{M})}_{\pm\sigma}(E))=&-\frac{J}{2}\zs\mean{S^z}^{(\text{M})}-\nonumber\\
&-\frac{J}{2}\zs\Gamma^{(\text{M})}_{\sigma}(E,G^{(\text{M})}_{\pm\sigma}(E))-\label{eqselfcpa}\\
&-\frac{J}{2}F^{(\text{M})}_{\sigma}(E,G^{(\text{M})}_{\pm\sigma}(E))\ .\nonumber
\end{align}
where $\mean{S^z}^{(\text{M})}$ is the magnetization of the impurity system. The subsystem's Green's function of component $\xi$ can be calculated via the projection\cite{velicky}
\begin{equation}
 G^{(\xi)}_{\sigma}(E)=\frac{\mean{G_{\sigma}(E)}}{1-(\eta^{(\xi)}_{\sigma}-\Sigma_{\sigma}(E))\mean{G_{\sigma}(E)}}\ .\label{eqproj}
\end{equation}
When we choose $\xi=M$ we get the (local) Green's function of the magnetic subsystem $G^{(\text{M})}_{\sigma}(E)$ and the system of equations is closed. The self-consistency cycle\footnote{We did not show the calculation of the expectation values and the chemical potential for readability. The expectation values have to be calculated with the projected functions $G_{\sigma}^{\text{(M)}}(E)$, $\Gamma_{\sigma}^{\text{(M)}}(E)$ and $F_{\sigma}^{\text{(M)}}(E)$ while the chemical potential is determined by the full Green's function $G_{\sigma}(E)$.} now reads 
\begin{align*}
 \Sigma_{\sigma}(E) &\stackrel{(\ref{eqgcpa})}{\longrightarrow} \mean{G_{\fk\sigma}(E)}\\
&\stackrel{(\ref{eqproj})}{\longrightarrow} G^{\text{(M)}}_{\sigma}(E)\\
& \stackrel{(\ref{eqselfcpa})}{\longrightarrow} M_{\sigma}(E)=\eta^{\text{M}}_{\sigma}\\
& \stackrel{(\ref{eq:cpa_selfconsistent})}{\longrightarrow} \Sigma_{\sigma}(E)\dots\ ,
\end{align*}
where the self-energy determination and the CPA formalism are strongly correlated.

\section{\label{secF}Free energy}
 
In the previous sections we determined the Green's function and the internal energy, but only for a given magnetization $\mean{S^z}$. To decide which magnetization the system actually favors we want to calculate the free energy and find its minimum, according to $\mean{S^z}$. To abbreviate the notation and avoid confusion with the entropy $S_M(T)$ we will use $M=\mean{S^z}$ for the local moment magnetization in this section. We note that $M$ is not the total magnetization $\tilde M = \mean{S^z}+\mean{\sigma^z}$ of the full system of local moments and electrons. But due to the self-consistent determination of $\mean{\sigma^z}(M) = n_{\uparrow}-n_{\downarrow}$ described in Sec. \ref{secGF}, $\tilde M$ is a unique function of $M$. Thus we can see the local moment's magnetization as the order parameter of the whole system.\\
The free energy is given by its general formula
\begin{align}
 F_M(T)=&U_M(T)-TS_M(T)\\
=&U_M(T)+T\dt F_M(T)\ .
\end{align}
Note that the magnetization is considered as a fixed (order-) parameter, i.e. an independent thermodynamic variable. Using the product rule and subtracting $U_M(0)=F_M(0)$ on each side we get
\begin{align}
-\frac{U_M(T)-U_M(0)}{T^2} = \dt\frac{F_M(T)-F_M(0)}{T}\ .
\end{align}
Integration of this equation with the condition
\begin{align}
 \lim_{T\rightarrow 0}\frac{F_M(T)-F_M(0)}{T}=-S_M(0)
\end{align}
 yields
\begin{align}
 F_M(T)=&U_M(0)-TS_M(0)-\label{eqFT}\\
&-T\underbrace{\int_0^TdT'\frac{U_M(T')-U_M(0)}{T'^2}}_{\equiv I_M(T)}\nonumber\ .
\end{align}
Since we know the internal energy $U_M(T)$ from the previous sections we can calculate the free energy if we have access to the entropy at $T=0$.\\
Equation (\ref{eqFT}) allows us to calculate the full temperature dependence of the free energy for a special magnetization. When we calculate it for many values of $M$ we get an ensemble of functions $\{F_M(T)\}$ from which we can get the magnetization dependence at a special temperature, $\{F_M(T)\}\rightarrow\tilde F_T(M)=\tilde U_0(M) - T (\tilde I_T(M)+\tilde S_0(M))$. If we find its minimum we get the magnetization at the temperature $T$:
\begin{align}
 0\stackrel{!}{=}&\left.\frac{\partial \tilde F_T(M')}{\partial M'}\right|_M\label{eqdFM}\\
=& \left.\frac{\partial (\tilde U_0(M') - T (\tilde I_T(M')+\tilde S_0(M'))}{\partial M'}\right|_M\nonumber\ .
\end{align}
Since this formula is exact, the quality of the results depend only on the accuracies of the internal energy and the entropy .

\subsection{Zero temperature entropy in the concentrated lattice}

We cannot determine the entropy in general so we make some approximations to find an expression for $S_M(0)$. The zero temperature entropy is defined as
\begin{equation}
 S_M(0)= k_B\ln\Gamma_M\ ,
\end{equation}
where $\Gamma_M$ is the number of ground states according to the magnetization $M$. First we assume that total number of states is given by the product of the number of states of the local moment and the itinerant electron system\footnote{This means that the number of states of one subsystem is independent of a special state of the other.}, $\Gamma_M=\Gamma^{\text{loc}}_M\Gamma^{\text{el}}_M$. This is plausible due to the different orders of magnitude of the dynamics of the local moments and the itinerant electrons. Thus the total entropy is the sum of the single entropies
\begin{equation}
 S_M(0) = S^{\text{loc}}_M(0)+S^{\text{el}}_M(0)\ .
\end{equation}
The assumption of separable electron/spin states is an ansatz which would be questionable, e.g., in systems where electrons are bound to distinct spins. For those cases other, more complex, methods to compute the entropy have to be found.\\
Let us start with the entropy of the local moments. Again due to the higher electron dynamics the local moments see no single static electron. Thus we can approximate the action of the electrons on the local moments as an effective field $B^{\text{eff}}_M$ which is equal for all lattice sites. This is a similar situation to the ideal paramagnet except that the field depends on electronic properties and the magnetization $M$ itself. We do not have to calculate $B^{\text{eff}}_M$ directly, but we can say that each configuration of local moments $\{m^z_i\}$ ($m_i^z=-S,\dots,S$) has the same energy as long as it has the same \emph{configurational} average $M'=\frac{1}{N}\sum_i m_i^z$. Just as well, configurations with different averages do have different energies. At $T=0$ only one energy level contributes to a thermodynamic average and therefore only states with the same configurational $M'$. So we can conclude that the thermodynamic magnetizations equals the configurational one, $M\stackrel{T=0}{=}M'$. Thus, for a given $M$, we have to count all possible configurations that have the same configurational average to get the zero temperature entropy. The identity of configurational and thermodynamical average is of course not true for $T>0$, but in this case we do not need it, because we determine the thermodynamical average by the minimization of the free energy. For $S=\halb$ we can give an analytical expression of the entropy
\begin{align}
 \Gamma^{\text{loc}}_M\stackrel{S=\halb}{=}&\frac{N!}{N_{\uparrow}!N_{\downarrow}!}\\
N=&N_{\uparrow}+N_{\downarrow},\quad M=\halb(N_{\uparrow}-N_{\downarrow})\nonumber\\
\frac{S_M(0)}{N}\stackrel{N\rightarrow\infty}{=}&-k_B\left((\halb-M)\ln(\halb-M)+\right.\nonumber\\
&\left.+(\halb+M)\ln(\halb+M)\right)
\end{align}
and for $S>\halb$ we have to count $ \Gamma^{\text{loc}}_M$ numerically.\\
The itinerant electrons see more or less a static potential of the local moments. That is why we use the expression of the Fermi gas
\begin{align}
 S_M^{\text{el}}(0)=&-k_B\sum_{\fk\sigma}\left[(1-\mean{n_{\fk\sigma}})\ln(1-\mean{n_{\fk\sigma}})+\right.\nonumber\\
&\left.+\mean{n_{\fk\sigma}}\ln \mean{n_{\fk\sigma}}\right]\label{eqS0el}
\end{align}
 to evaluate the zero-temperature entropy. Note that the mean values $\mean{n_{\fk\sigma}}$ are calculated by the full theory of section \ref{secGF}!

\subsection{\label{secS0dil}Zero temperature entropy in the diluted lattice}

We need the same simplifications as in the previous section. Therefore nothing changes in the formula for the electronic entropy (\ref{eqS0el}), except that we get $\mean{n_{\fk\sigma}}$ from the CPA Green's function.\\
For the local moments we define the magnetic subsystem's magnetization at $T=0$ as $M=\frac{1}{N_M}\sum_i'm_i^z$ with $N_M=xN$. When we still assume that the local moments interact only with an effective field created by the electrons (corresponding to an ideal paramagnet or independent spins, respectively), we do not have to change the general procedure to get $S_0(M)$. Actually the entropy per lattice site stays the same as in the concentrated case. Thus we get
\begin{align}
 \frac{S_M(0,x=1)}{N} =& \frac{S_M(0,x)}{N_M}\nonumber\\
\Rightarrow S_M(0,x)=&xS_M(0,x=1)\ .
\end{align}
Note that the disorder itself leads to no (magnetization dependent) contribution to the entropy in this case, but only the dilution. Thus special disorder effects on the magnetization, e.g. found in Ref.\cite{berciu}, are beyond the scope of this paper. They could, e.g., play a role in materials which tend to form clusters.\\
It is a big advantage that we do not need to map the KLM onto a Heisenberg model to calculate the local moment's properties, especially in the diluted system. This would involve extra disorder\cite{tang,bouzrkky,bouzproj} within the Heisenberg model to be calculated separately from the electronic disorder already kept within the CPA formalism. Just as the general inconsistency of treating one system with two models (KLM, Heisenberg) this problem is avoided in our treatment.
Since our theory is based on translational invariance within each component we do not get percolation effects. Those can be crucial at low doping if there is a short range interaction between the local moments. But even for an interaction range slightly larger than next-neighbor, the percolation threshold is strongly reduced\cite{hilbert}. Since the interaction between local moments in the KLM is assumed to be RKKY-like and therefore long ranged, percolation effects should play a minor role.\\
As mentioned before, in systems which are not translational invariant, the entropy had to be calculated more appropriate, which would be a difficult task indeed.

\section{Results}
\begin{figure}[tb]
 \includegraphics[width=.9\linewidth]{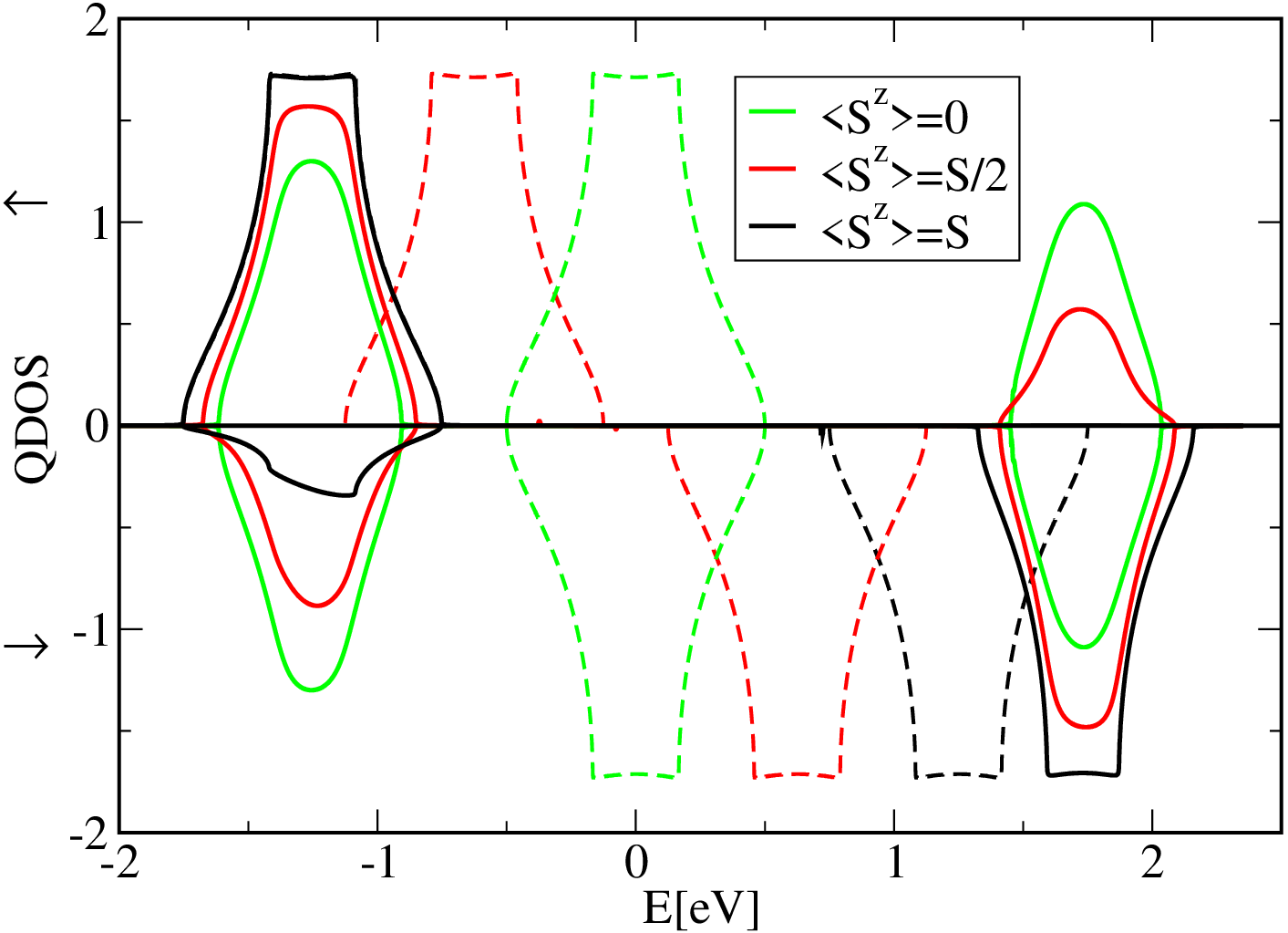}
\caption{\label{figqdosvarSz}(color online) The quasi-particle density of states of the concentrated system for various magnetizations $\mean{S^z}$. The QDOS splits into two subband at $E\approx-\halb JS$ and $E\approx \halb J(S+1)$. For decreasing $\mean{S^z}$ spectral weight is transferred from upper to the lower subband or vice versa (MCDA). The mean-field QDOS (dashed lines) is only the shifted (by $\zs\halb J\mean{S^z}$) free QDOS. Which magnetization $\mean{S^z}$ is actually present for a specific parameter set, has to be calculated by the minimization of the free energy. Parameters: $S=\frac{5}{2}$, $W=1$eV, $J=1$eV, $n=0.05$}
\end{figure}
Our main interest is focussed on the dependence of the Curie temperature on the significant model parameters. That's why we first want to discuss these dependencies in the concentrated ($x=1$) system. After that we come to the diluted one. Finally we want to use model parameters which are suitable for the real system of Ga$_{1-x}$Mn$_x$As to compare our theoretical results with those of the experiment.\\
It turns out that in the KLM the term $\partial_M\tilde I_T(M)$ in Eq. (\ref{eqFT}) is much smaller than $\partial_M \tilde S_T(M)$. Thus we can neglect this in a very good approximation. Since $\partial_M\tilde I_T(M)$ was the only term with an explicit temperature dependence we get a simplified formula
\begin{align}
 T(M)=\left.\frac{\partial_{M'}\tilde U_0(M')}{\partial_{M'}\tilde S_0(M')}\right|_M\label{eqTM}
\end{align}
and its inverse function $M(T)$. It means that the magnetization curve is only dependent on zero temperature quantities\footnote{Zero temperature does not mean the absolute ground state in our work, since we calculate $U$ for different magnetizations, which are not equal in energy}. Especially we can use (\ref{eqTM}) to calculate the Curie temperature, when we set $M=0^+$.\\
The lattice is assumed to be simple cubic in this model study with a next-neighbor tight-binding dispersion
\begin{align}
 \ek=\frac{W}{6}(\cos(k_xa)+\cos(k_ya)+\cos(k_za))
\end{align}
where $W$ is the free bandwidth.

\subsection{\label{secrescs}Concentrated System}

To calculate $T_C$ the quantity of interest is the internal energy, in our case. Thus we should have a look on the quasi-particle density of states (QDOS) $\rho_{\sigma}(E)=-1/(\pi N)\sum_{\fk}\text{Im}G_{\fk\sigma}(E)$, since it has the main influence on $U$ as it can be seen in Eq. (\ref{eqU}). In the concentrated system the QDOS splits into two subbands at different energies (cf. Fig. \ref{figqdosvarSz}). One is situated at $E\approx -\frac{J}{2}S$ (lower subband for $J>0$ and upper for $J<0$) and the other at $E\approx \frac{J}{2}(S+1)$. In contrast to the MF approximation these subbands remain at this energy range for all values of $\mean{S^z}$, only the spectral weight is shifted from the upper to the lower subband for decreasing magnetization, or vice versa. This means a much lower energy difference $\Delta U =U(\mean{S^z}+d\mean{S^z})-U(\mean{S^z})$ in the MCDA than in the MF approximation. Comparing this with Eq. (\ref{eqTM}) shows that the MCDA results should give lower Curie temperatures as the MF approximation.
\begin{figure}[tb]
 \includegraphics[width=.9\linewidth]{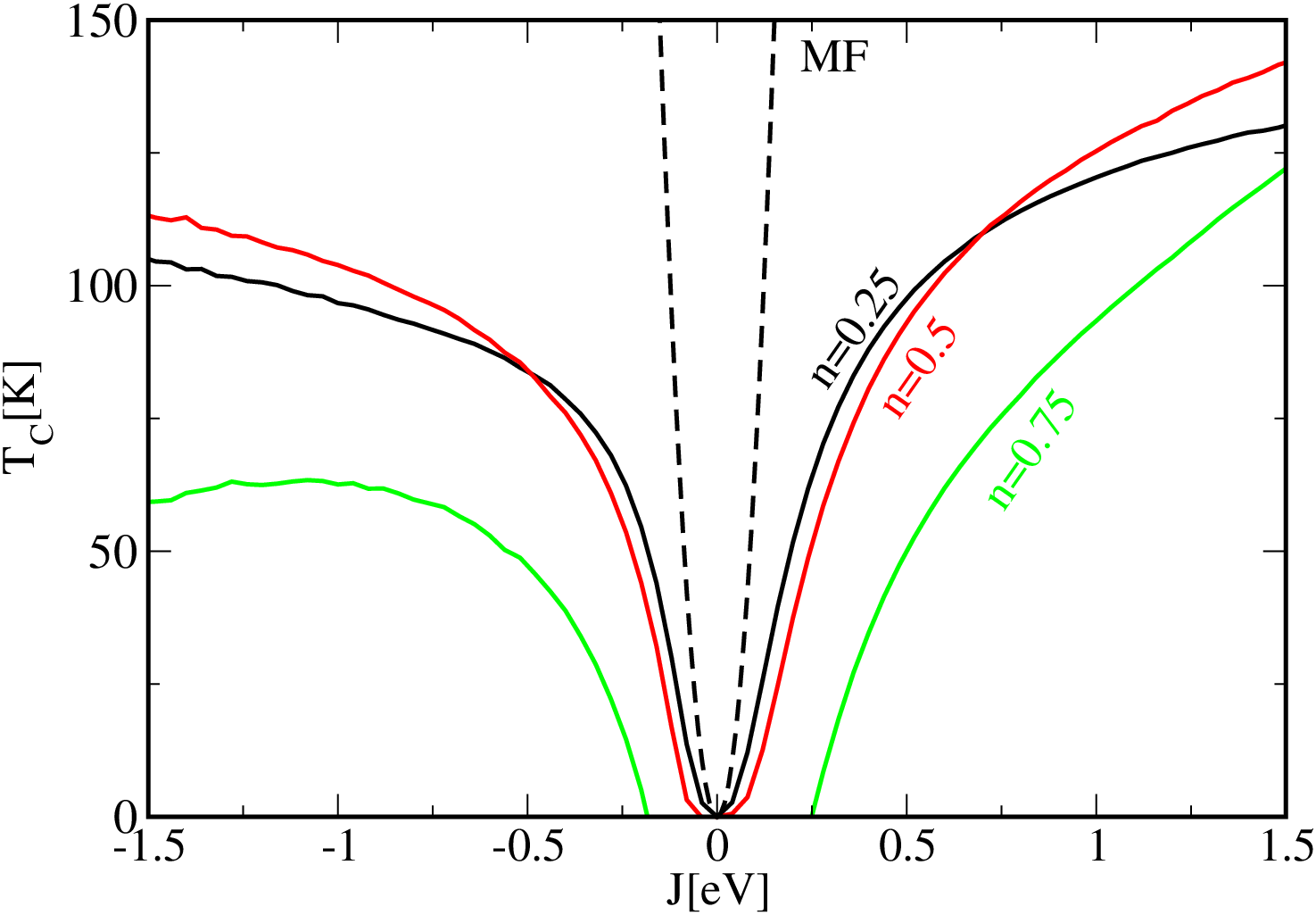}
\caption{\label{figTcJ}(color online) Curie temperature vs. coupling $J$ for various electron densities $n$. In contrast to the mean-field result $T_C$ saturates for large $|J|$ or is growing slowly, respectively. At the MF approximation (dashed line, $n=0.05$) $T_c$ increases quadratically with $J$. Parameters: $S=\frac{5}{2}$, $W=1$eV}
\end{figure}
Indeed this is the case, e.g. when we plot $T_c$ over $J$ (Fig. \ref{figTcJ}). The MF Curie temperatures are much higher\footnote{Because of the high $T_Cs$ of the MFA we usually use a smaller $|J|$ for MF results in this work to have a better comparison to the MCDA values.} than those of the MCDA. Additionally we find that $T_C^{\text{MF}}\sim J^2$ while we get saturation or only a slight change at large $|J|$ within the MCDA.\\
\begin{figure}[tb]
 \includegraphics[width=.9\linewidth]{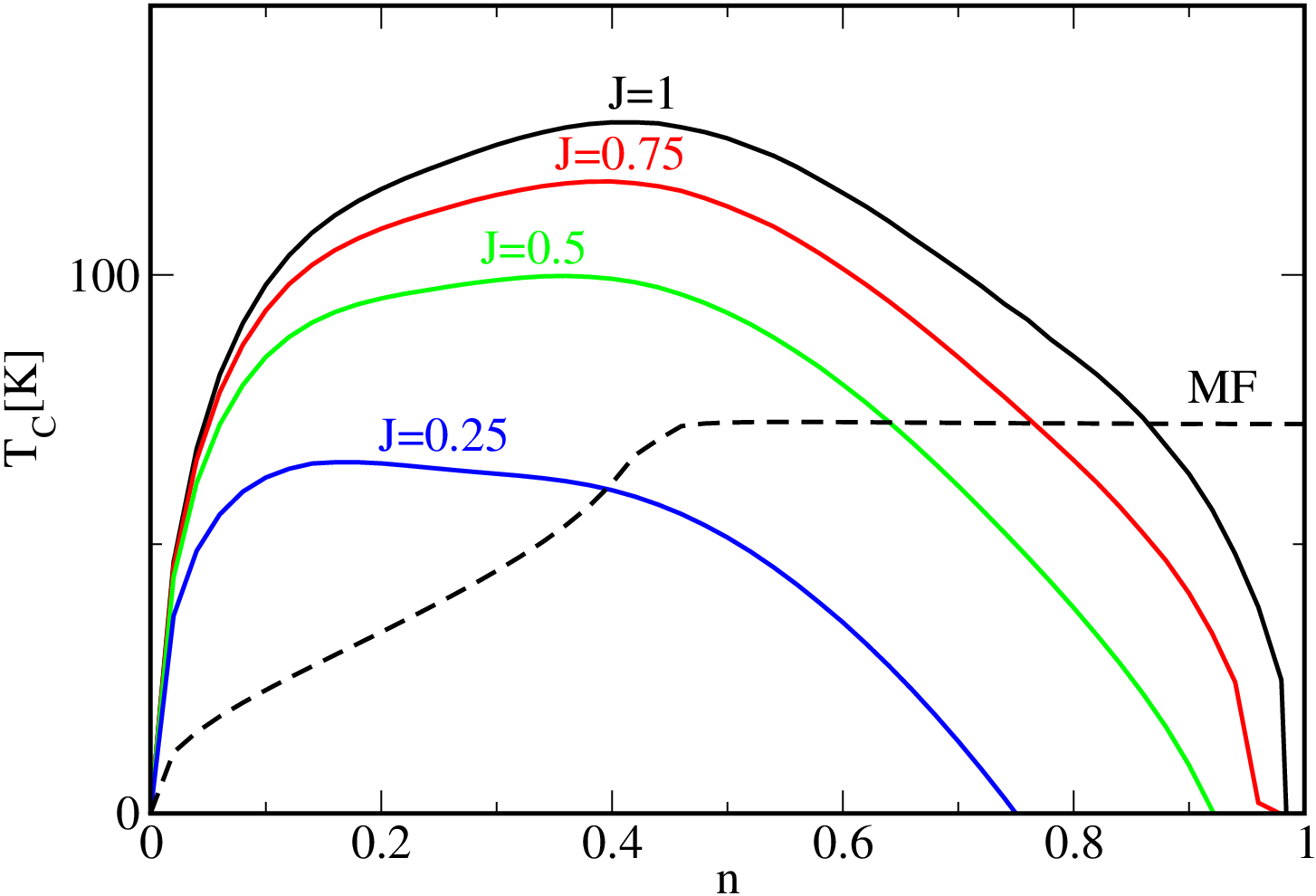}
\caption{\label{figTcn}(color online) Curie temperature vs. electron density $n$ for various couplings $J$. The Curie temperature of the MCDA vanishes at half-filling ($n=1$) even for large $J$ contrary to the mean-field approximation (dashed line, $J=0.05$). Parameters: $S=\frac{5}{2}$, $W=1$eV}
\end{figure}
The next important parameter is the electron density $n$ or the hole density $p=2-n$ in the conduction band. In the MFA, due to the shift of the majority spin's QDOS by $-\frac{|J|}{2}\mean{S^z}$, the energy difference $\Delta U$ is always positive, also resulting in a positive $T_C$ for all $n, p$ (one can also see this direct connection in the shape of the $T_C$-$n$-curve in Fig. \ref{figTcn} which resembles the shape of the free QDOS). Contrary to that the MCDA has always a vanishing $T_C$ at half-filling (cf. Fig. \ref{figTcn}). Similar results can be found in other publications\cite{mcda1997,tang,mRKKYW}. This is a very crucial result, contrary to a MF approximation which shows a maximum $T_Cs$ at half-filling (dashed line in Fig. \ref{figTcn}). The reason for this behavior is that for $n\rightarrow 1$ the chemical potential $\mu$ is at the upper band edge of the lower subband. As Fig. \ref{figqdosvarSz} shows this upper band edge is shifted to lower energies for decreasing $\mean{S^z}$ within our theory which means a reduction of the internal energy. Thus the energy difference $\partial_M U_0(M)$ in Eq. (\ref{eqTM}) gets negative resulting in a negative $T_C$, being thus non-physical.\\
\begin{figure}[tb]
 \includegraphics[width=.9\linewidth]{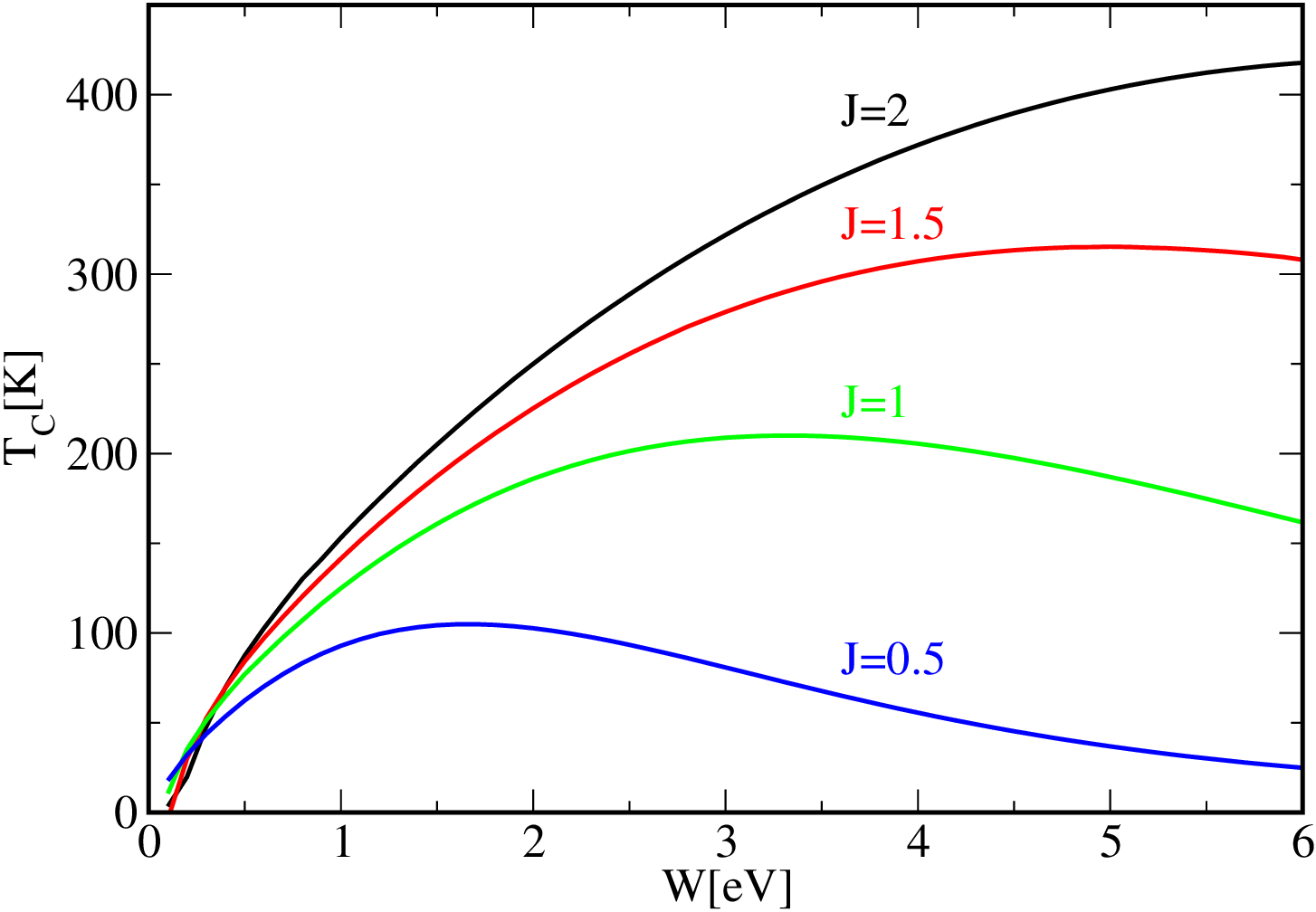}
\caption{\label{figTcW}(color online) Curie temperature vs. free bandwidth $W$ for various couplings $J$. In the strong coupling limit ($JS\gtrsim W$) $T_C$ increases approximately linearly with $W$. At lower couplings the effect is inverse and an increase of $W$ leads to a reduction of $T_C$. Parameters: $S=\frac{5}{2}$, $n=0.5$}
\end{figure}
As the third parameter remains the free bandwidth $W$. It is known\cite{mRKKYW} that in the strong coupling limit ($JS\gtrsim W$, the subbands are split) $T_C$ becomes independent of $J$ (double exchange limit). In this case the Curie temperature is proportional to $W$ as can be seen in Fig. \ref{figTcW}. But at lower $J$ an inverse effect occurs and an increase of $W$ leads to a reduction of the Curie temperature $T_C\sim JS/W$ because the bands are not split any more. This is the typical RKKY behavior which is valid at low couplings.\\
We now can determine the optimal parameters for high Curie temperatures in the KLM. $T_C$ increases with the magnitude of the coupling $|J|$. But for large $|J|$ saturation of $T_c$ occurs and the increase is limited. The electron (hole) density should be at quarter-filling for strong couplings or lower for a smaller $|J|$. Finally the free bandwidth should be preferably large as long as system stays in the strong coupling regime ($\frac{JS}{W}>1$).\\
Indeed the findings in this section are similar to studies done with a mapping of the KLM onto an Heisenberg model\cite{carlos2002,michev,michev2,stier}. Also in these approaches, which go beyond mean-field, a saturation of $T_C$ with increasing $|J|$ and a vanishing of ferromagnetism below/at half-filling have been found. Thus both methods support each other. As stated before (Sec. \ref{secS0dil}) the use of the free energy has some advantages for computing the magnetization of diluted system. Of course one can get additional information from the free energy itself, like the entropy or heat capacity. It also can be useful to compare free energies of different magnetic phase to make statements about their existence at finite temperatures. This will be left for future work. 

\subsection{\label{secresds}Diluted system}
\begin{figure}[tb]
 \includegraphics[width=.9\linewidth]{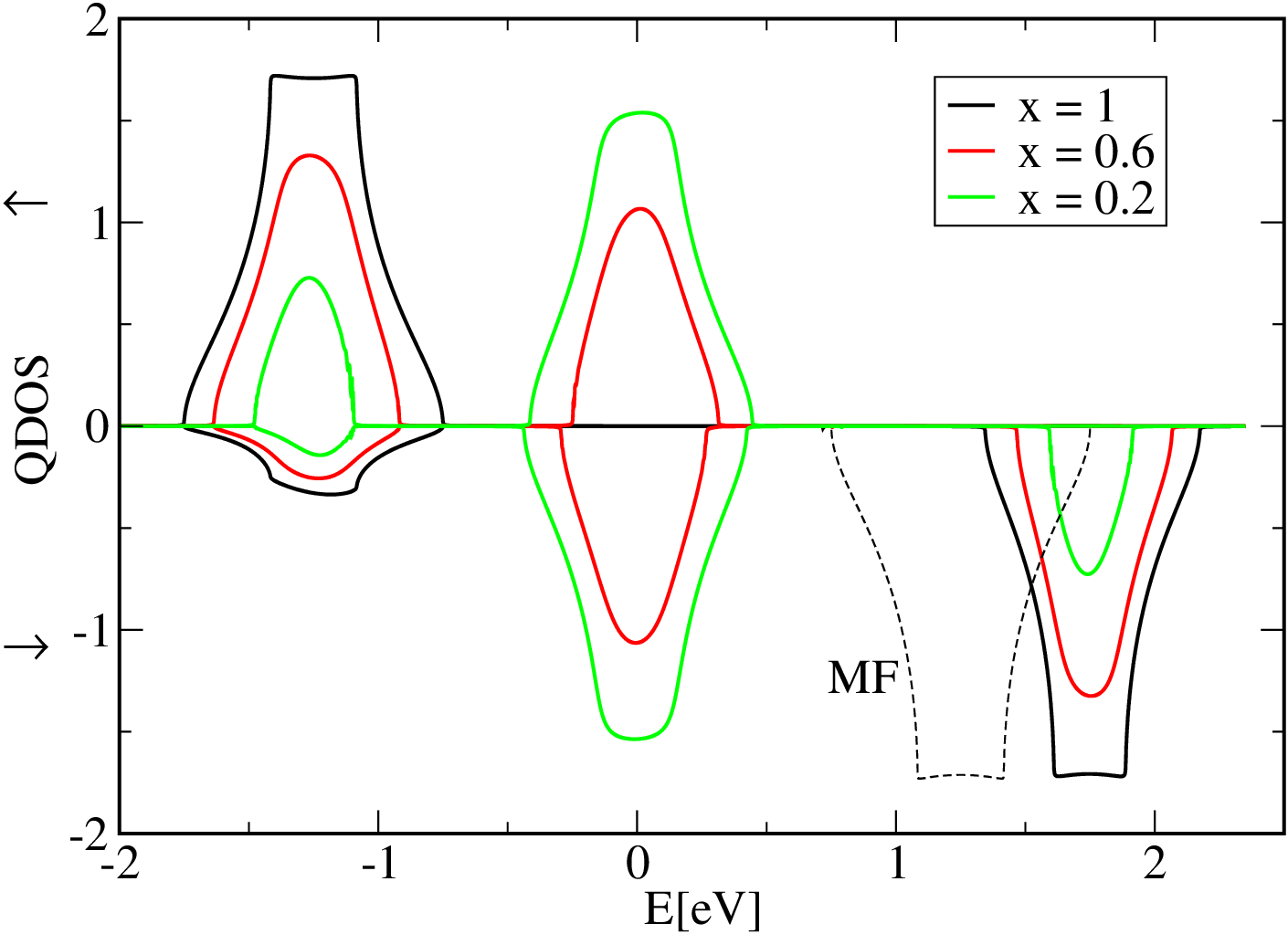}
\caption{\label{figqdosvarx}(color online) The quasi-particle density of states of the carriers for various dilutions $x$. The correlated sub-bands around $E\approx-\halb JS$ and $E\approx \halb J(S+1)$ arise from the interaction of the carries with the localized spins $\fS_i$ and have a total spectral weight of $2x$. Around $E\approx  T_0^{\text{NM}}=0$ a non-correlated sub-band exist resulting from the lattice sites without magnetic impurities. Note that the ferromagnetically saturated state, which is shown here, is not necessarily thermodynamically stable for all carrier densities $n$ and or couplings $J$. Parameters: $S=\frac{5}{2}$, $W=1$eV, $J=1$eV, $T_0^{\text{NM}}=0$, $n=0.01$, $\mean{S^z}=S$}
\end{figure}
The dependencies of $T_C$ on the model parameters in the diluted system stay approximately the same as in the concentrated one. But there are some additional effects due to the dilution.\\
In diluted systems the carriers are on lattice sites with or without magnetic impurities. Due to the interaction with the localized spins $\fS_i$ two distinct parts of the QDOS of the carriers arise (Fig. 5). One is located around $E\approx 0$ (or $T^{\text{NM}}$, resp.) and results from the non-impurity sites (non-correlated part). The two other sub-bands are at $E\approx -\halb JS$ and $E\approx \halb J(S+1)$ at the same positions as the quasi-particle sub-bands of the concentrated KLM (correlated part).\footnote{Note that Fig. 5 shows, for graphical reasons, a scenario of strong interaction $JS/W$ and all parts of the QDOS belong to the itinerant carriers' band (valence band). The correlated sub-bands should not be confused with the $d$-bands of the DMS which are taken into account by the local moments in our model. The splitting in real DMS would be much smaller.}. The positions of the sub-bands are not affected by the concentration $x$, but the spectral weights are. In Fig. \ref{figqdosvarx} one sees that the correlated subbands have the spectral weight $2x$ and the non-correlated $2(1-x)$. This means, as long as the correlated bands are split from the non-correlated band, that we get an effective filling $n^{\text{eff}}=n/x$ of the correlated band and, e.g., half-filling of the correlated bands occurs at $x<n<2-x$.
\begin{figure}[tb]
 \includegraphics[width=.9\linewidth]{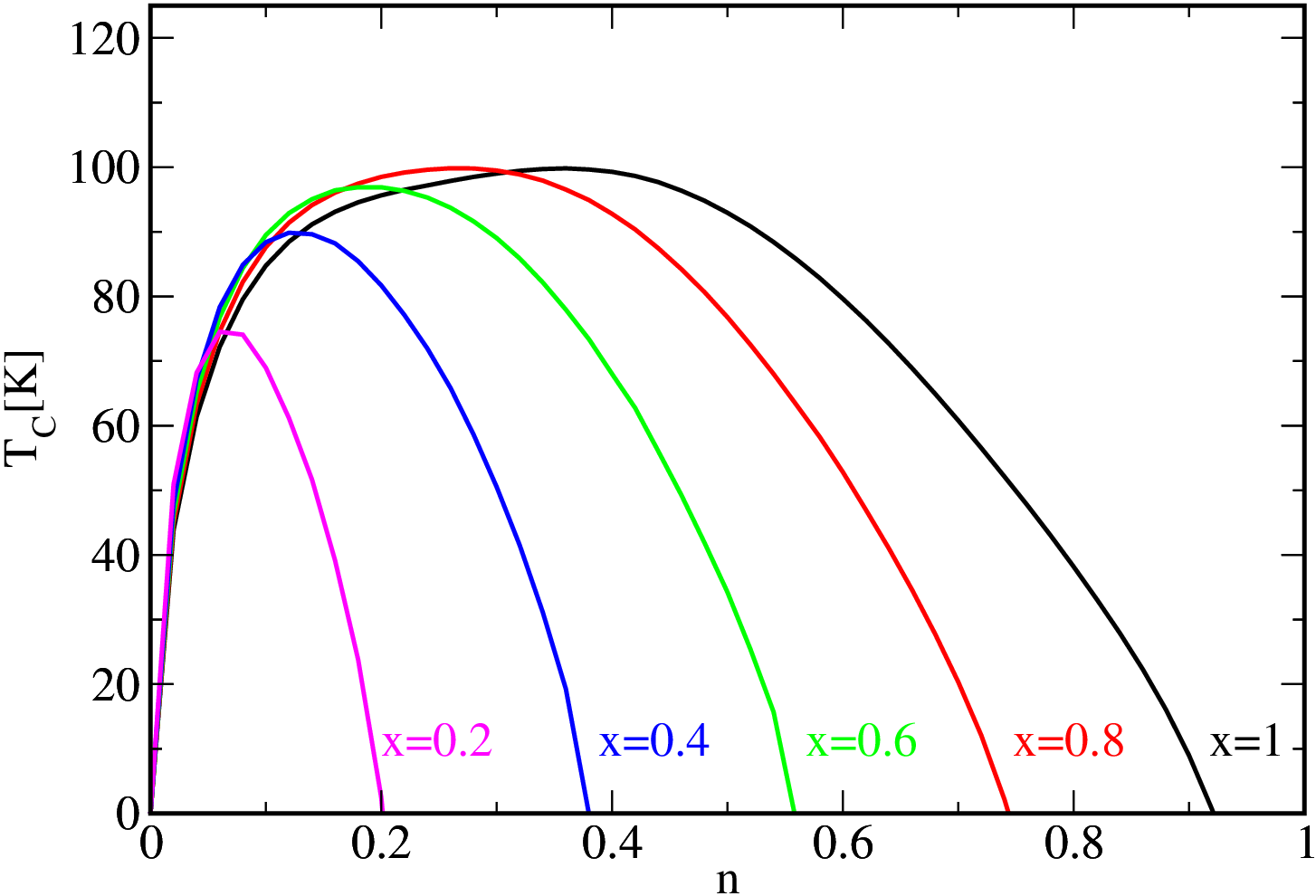}
\caption{\label{figTcnvarx}(color online) Curie temperature vs. electron density $n$ for various dilutions $x$. A finite positive Curie temperature only occurs below half-filling of the correlated band ($n<x$). Parameters: $S=\frac{5}{2}$, $W=1$eV, $J=0.5$eV, $T_0^{\text{NM}}=0$} 
\end{figure}
As seen in the case of the concentrated KLM ($x=1$) ferromagnetism at half-filling of the sub-bands is not thermodynamically stable. Thus there are positive Curie temperatures only for $n<x$ (partly filled lower correlated subband, cf. Fig. \ref{figTcnvarx}) or $n>2-x$ (partly filled upper correlated subband) . For $x<n<2-x$ the chemical potential $\mu$ is in the non-correlated band (for $T^{\text{NM}}_0=0$ at least) andthe lower/upper correlated subband is completely filled/empty resulting in a vanishing of ferromagnetism.\\
Secondly the maximum $T_C$ at quarter-filling is reduced for decreasing $x$. It can be understood, in the strong coupling regime, with the reduction of the effective bandwidth of the correlated band\cite{sato}, $W^{\text{eff}}\approx \sqrt{x} W$. Since we know that $T_C$ is proportional to $W$ for large $|J|$ the Curie temperature should be $T_C(n^{\text{eff}})\approx \sqrt{x}T_C(n,x=1)$ in the diluted system. Note that the strong coupling regime is now defined by the effective bandwidth, $JS\gtrsim W^{\text{eff}}$. Thus with decreasing $x$ a system can get into the strong coupling limit for constant $J$.\\
One of the best known DMS is Ga$_{1-x}$Mn$_x$As. To model this material we fix some parameters from now on. As seen from photoemission experiments\cite{photo} the coupling is about $J\approx-1$eV. Furthermore we approximate the bandwidth from ab-initio calculations\cite{gmaw4} to $W=4$eV. Since we use a simple cubic tight-binding dispersion we take into account only the main part of the "ab-initio"-DOS below/around the fermi-level for this estimation. Furthermore it is useful to change to the hole picture, where we define the hole density $p=2-n$ since the bands are completely filled for $n=2$.\\
Of course, in a real material other mechanisms besides magnetic interactions play a crucial role. For example the atomic levels of different atoms will be different in general. To take those effects into account we can change the position of the non-correlated band by choosing a specific center of gravity $T^{\text{NM}}_0$, which we have introduced in the model Hamiltonian (\ref{eqH2}). Depending on the value of $T^{\text{NM}}_0$ the magnetic subbands can be split from the correlated band or can lie within it. Especially this determines which kind of holes are created by filling the band with carriers. When the upper magnetic subband lies above the non-correlated band, holes preferentially occupy magnetic sites for low doping leading to a large "magnetic hole density" $p^{\text{M}}$ (cf. Fig. \ref{figqdosgamnas}). This will be reversed when the magnetic subband gets to lower energies.
\begin{figure}[tb]
 \includegraphics[width=.9\linewidth]{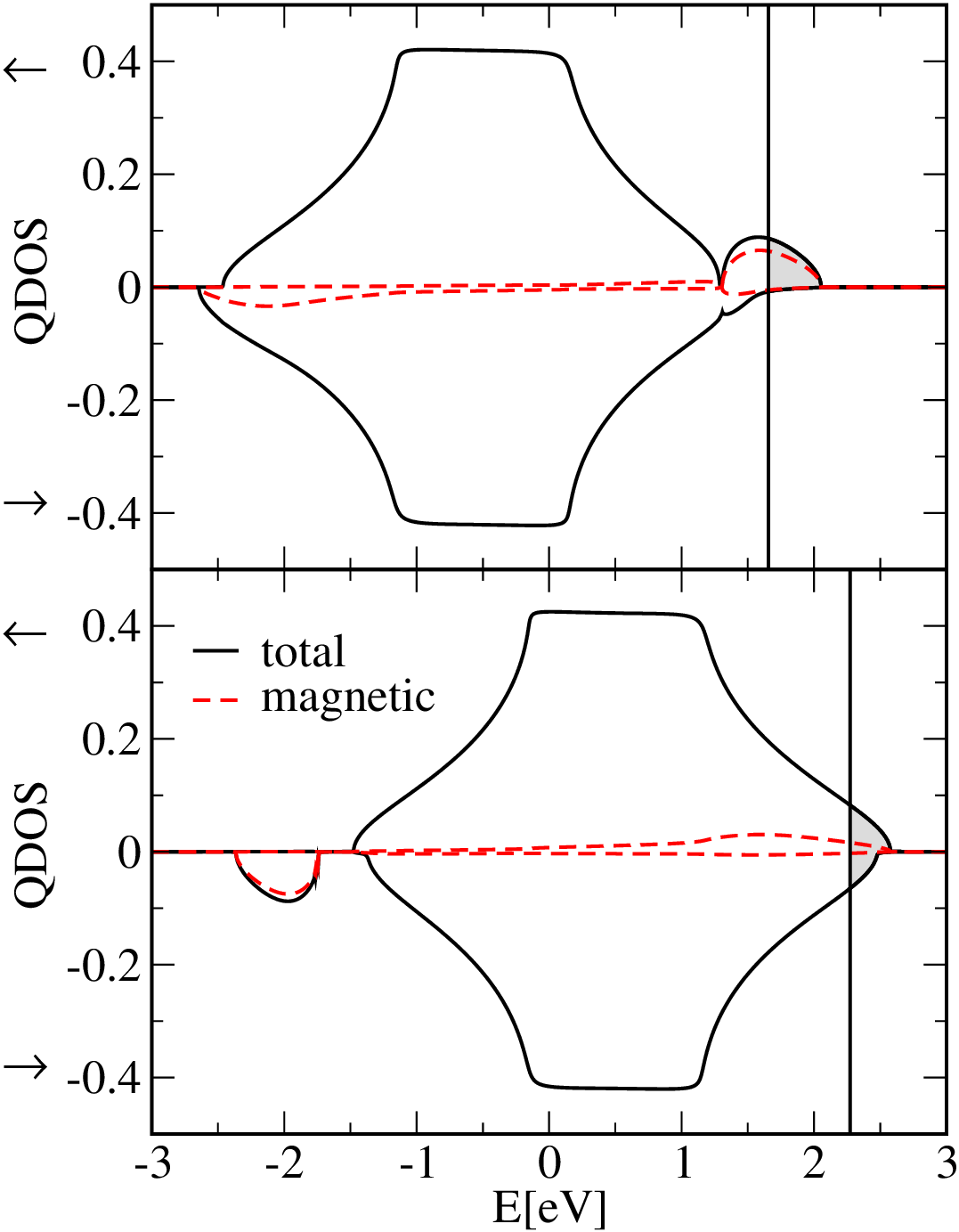}
\caption{\label{figqdosgamnas}(color online) Electronic QDOS of the diluted system for two different centers of gravity of the noncorrelated band $T^{\text{NM}}_0$. The total QDOS (black line), the projected QDOS of the magnetic sites (red dashed line) and the Fermi level (black vertical line) are shown. \emph{Top}: $T^{\text{NM}}_0=-0.5$eV. The upper subband stemming from the magnetic sites  is slightly above the noncorrelated band and the lower one is broadened within the noncorrelated band. Holes exist above the Fermi level (grey area) almost completely in the magnetic QDOS leading to a ratio $p^{\text{M}}/x=0.357$ (ratio for all holes $p/x=0.5$) . \emph{Bottom}: $T^{\text{NM}}_0=0.5$eV. The upper impurities' subband is within the noncorrelated band. The number of holes in the magnetic QDOS is reduced to $p^{\text{M}}/x=0.07$ .  Parameters: $S=\frac{5}{2}$, $W=4$eV, $J=-1$eV, $x=0.05$, $p=0.025$, $\mean{S^z}=S$, $T=0$K}
\end{figure}
The reduction of $p^{\text{M}}$ with increasing $T^{\text{NM}}_0$ has two different effects on $T_C$. For low carrier densities (left of maxima of the parabola-like curves in Fig. \ref{figTcnvarx}) a decreasing $p^{\text{M}}$ means a reduction of $T_C$. On the other hand, systems with $p>x$ reach a finite Curie temperature not before a certain amount of holes on magnetic sites is transferred to nonmagnetic sites (cf. Fig. \ref{figTcTNM}).
\begin{figure}[tb]
 \includegraphics[width=.9\linewidth]{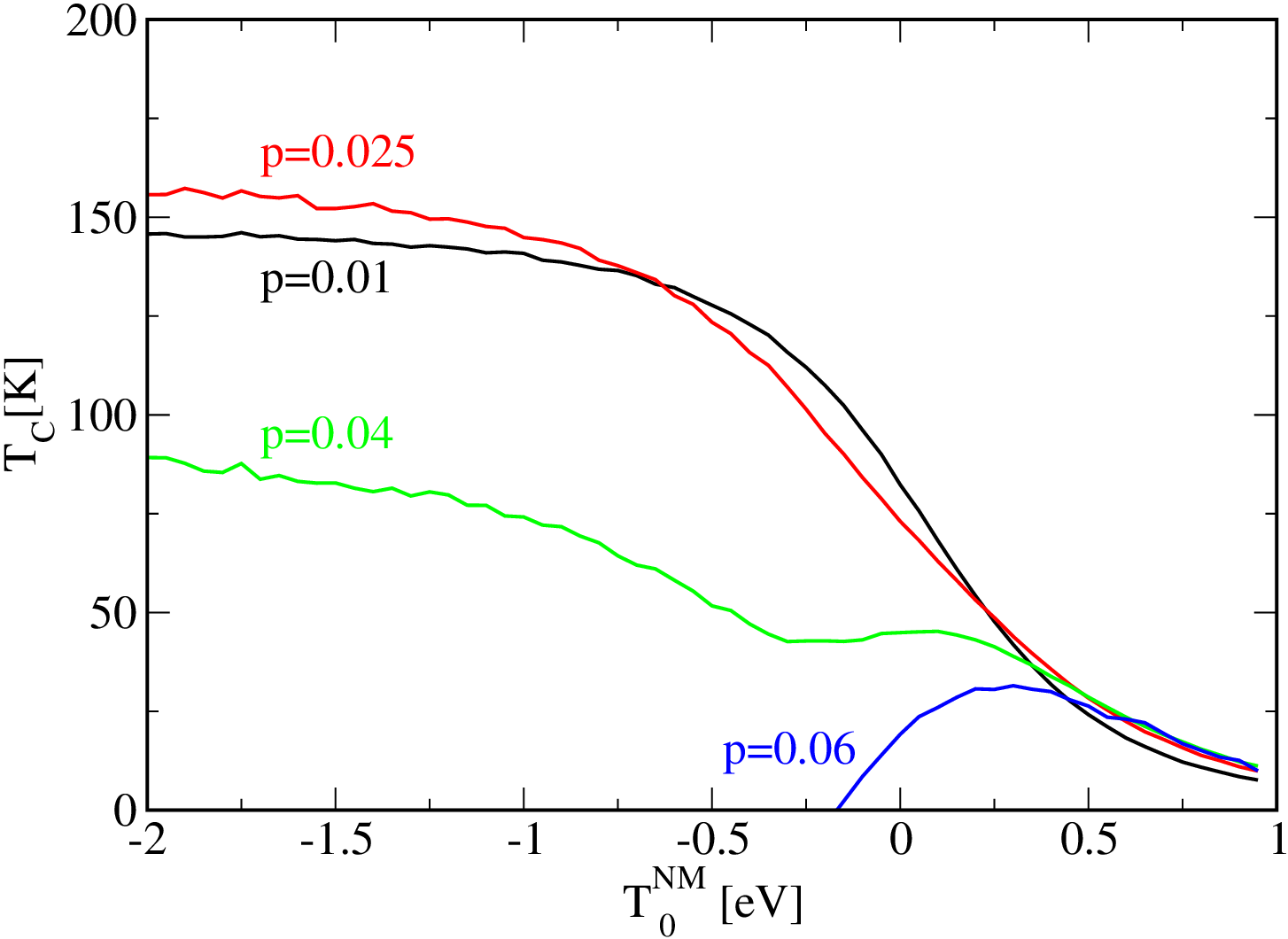}
\caption{\label{figTcTNM}(color online) Curie temperature vs. the center of gravity of the noncorrelated band $T_0^{\text{NM}}$ for various hole densities $p=2-n$. As long as chemical potential is in the correlated band $T_c$ is large. When the correlated band (and therewith $\mu$) gets into the noncorrelated band (here at $T^{\text{NM}}_0\approx (-0.5\dots 0.5)$eV) $T_c$ decreases for low hole densities. For hole densities $p>x$ finite Curie temperatures occur only in the non-split regime. Parameters $S=\frac{5}{2}$, $W=4$eV, $J=-1$eV, $x=0.05$}
\end{figure}
Figure \ref{figTcp} shows clearly the dependence of $T_C$ on the filling of the magnetic subband. As long as $p^M/x$ does not reach a critical value a finite Curie temperature exists. For the case that the correlated subband is far above the non-correlated band and a large $|J|$ this critical filling would be approximately one. If the subband comes near (but not into) the non-correlated band hybridization effects with the non-correlated band become larger and a small amount of spectral weight is transferred to the energy region of the non-correlated band. Thus the spectral weight of the correlated subband outside the non-correlated band decreases below $x$. Nevertheless the complete actual filling in Fig. \ref{figTcp} can be connected to the kinks in the $p^{\text{M}}-p$ curves which are at the same position where $T_C$ reaches zero. These kinks disappear when the upper correlated subband gets into the non-correlated band. This means that a complete filling of the subband cannot be achieved even at hole densities $p\approx x$, but at much higher hole densities $p\gg x$ (cf. Fig. \ref{figqdosgamnas}). Thus the Curie temperature stays above zero for $p>x$. Even though in this case ferromagnetism occurs over a broad range of the hole density, the maximum $T_C$ is heavily reduced. A very similar picture has been found for dynamical mean-field/Monte Carlo calculations\cite{popescu}.\\
\begin{figure}[tb]
 \includegraphics[width=.9\linewidth]{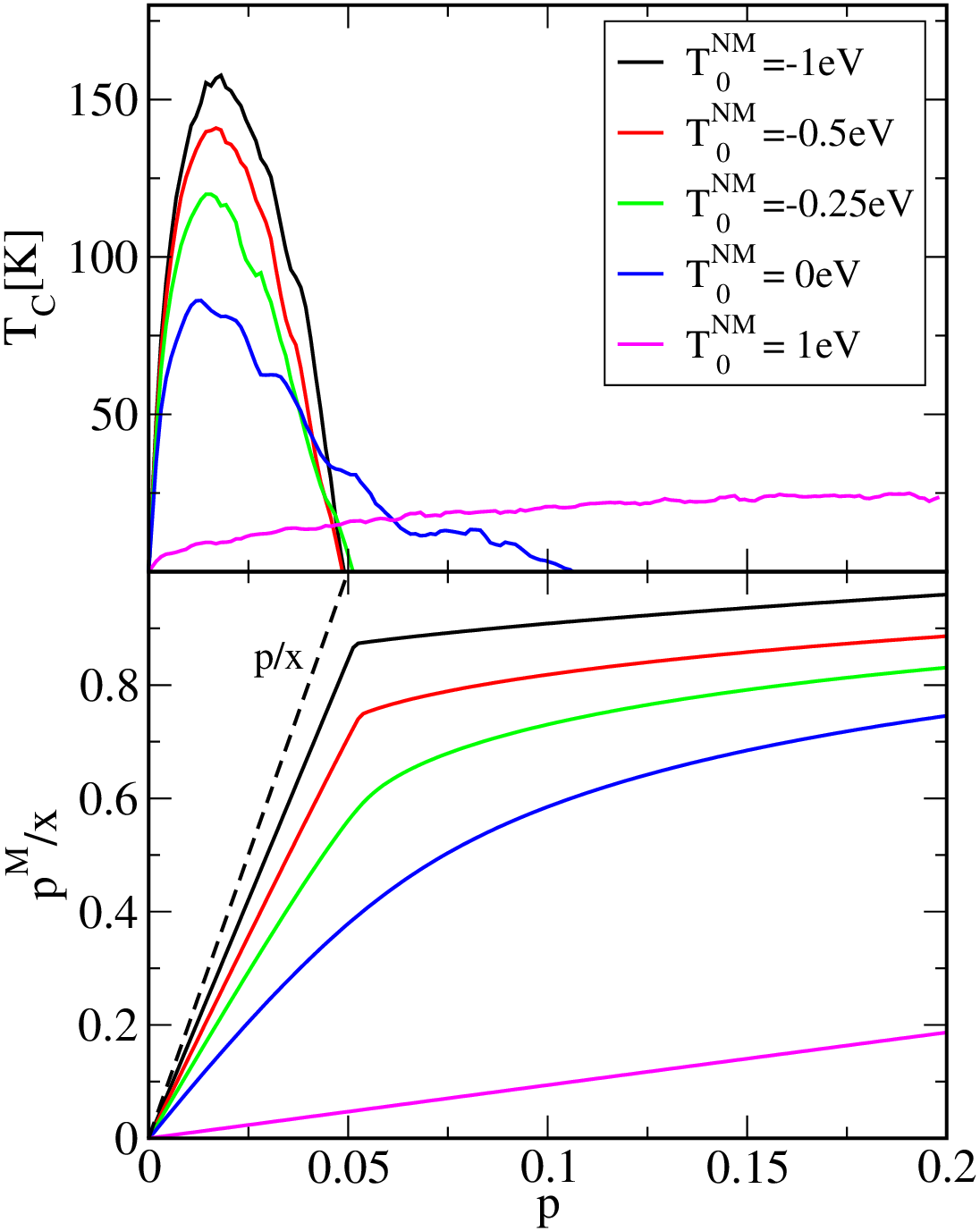}
\caption{\label{figTcp}(color online) \emph{Top}: Curie temperatures vs. hole density $p=2-n$ for different centers of gravity $T^{\text{NM}}_0$. \emph{Bottom}: The according hole density on the magnetic sites $p^{\text{M}}/x$. $T_C$ is heavily affected by the filling of the magnetic subband (cf. Figs. \ref{figqdosgamnas}, \ref{figTcTNM}). When the upper correlated band is split from the non-correlated band ($T^{\text{NM}}_0 \lesssim -0.25$eV),  $p^{\text{M}}/x$ increases rapidly with $p$. In this case a distinct critical ratio can be found where ferromagnetism breaks down ($T_C=0$). If the subband is within the correlated band there is ferromagnetism over a broad range of hole densities but the maximum $T_C$ is reduced. Parameters: $S=\frac{5}{2}$, $W=4$eV, $J=-1$eV, $x=0.05$}
\end{figure}
In fact ab-initio calculations predict that some amount of states, arising from the doping with manganese, is slightly above the non-correlated band\cite{gmaw4}. This is in agreement with measurements of \emph{Cho et al.} \cite{expcomp,expcomp2} on the behavior of ferromagnetism. They state that for low hole density an increase of $p$ leads to an increase of $T_C$. Beyond a specific $p$ $T_C$ decreases and finally breaks down as calculated in this work in the "split-up" regime. Unfortunately \emph{Cho et al.} cannot give absolute values of $p$. Why do experiments and ab-initio calculations show a maximum $T_C$ at $p\approx x$ as found by \emph{Jungwirth et al.}\cite{dmsMF} in contrast to our model theory? One reason can be that the correlated bands are partly within the non-correlated band which allows a finite $T_C$ even at $p/x\approx 1$ , because the correlated band is \emph{not} completely filled in this case. But more important is that the $pd$-levels of the impurities are threefold degenerated. One hole in these three states mean effectively a "third filling" of the correlated band\cite{bouzuni}. Thus even for $p=x$ the correlated band is not completely filled, but on the left side of the maxima of the parabolas in Fig. \ref{figTcp} with an effective filling of $p^{\text{eff}}\approx \frac{1}{3}$.  This would indeed explain the increase of Curie temperatures with increasing hole density and the breakdown of ferromagnetism at large $p$.\\
Since we have seen that highest Curie temperatures occur in the "split-up" regime and ab-initio calculations also predict this scenario we choose a $T^{\text{NM}}_0$ where the bands are split to investigate the influence of manganese doping on $T_C$. In this case an only partly filled correlated band is necessary for finite $T_C$s. As the real variation of the hole density with doping is not completely clear we propose three different types of fillings/compensations in Fig. \ref{figTcx}.
\begin{figure}[tb]
 \includegraphics[width=.9\linewidth]{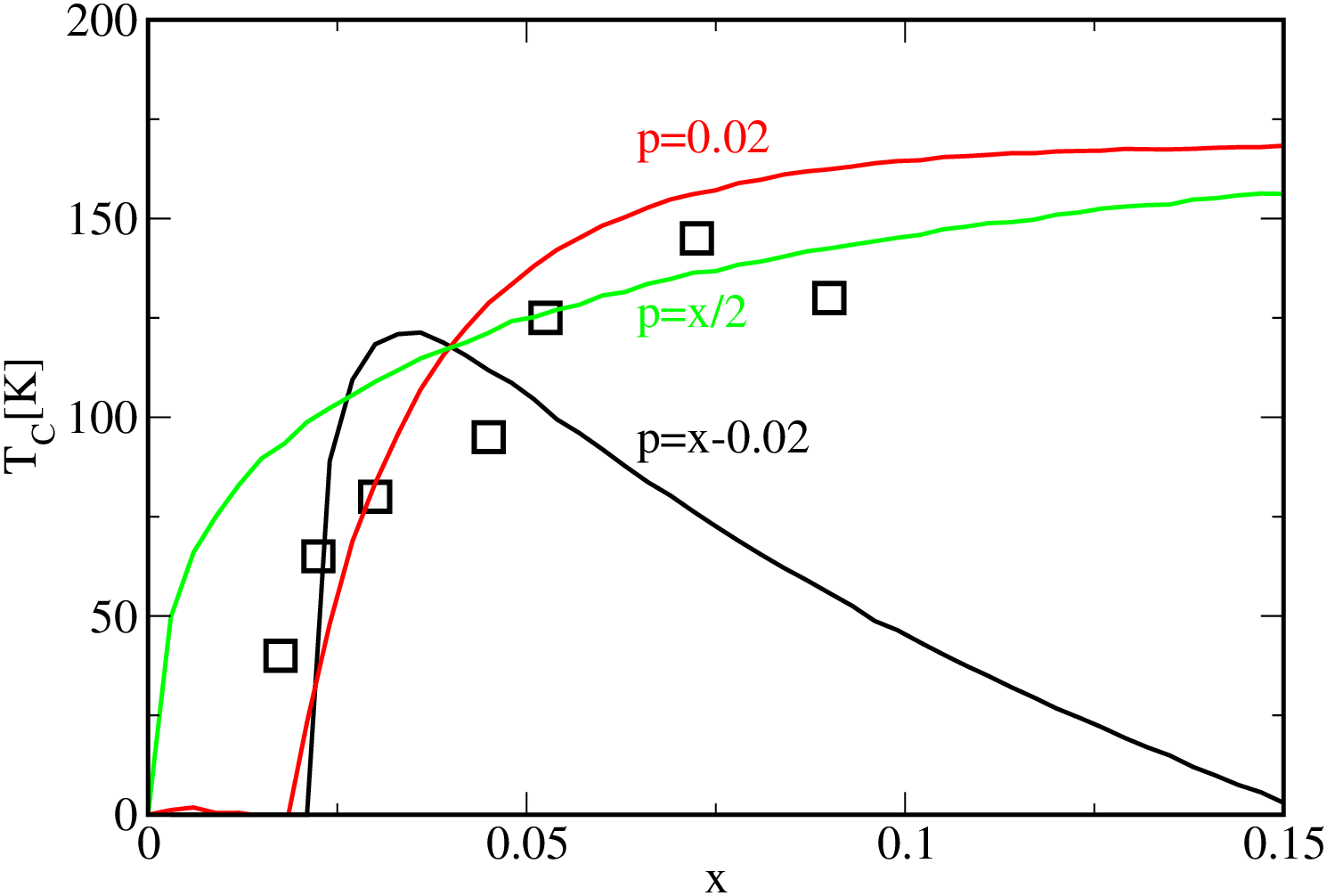}
\caption{\label{figTcx}(color online) Curie temperature vs. doping rate $x$. Three curves are plotted for different compensation types. A constant hole density (red line, $p=0.02$) or compensation $c$ (black line,$p=x-c=x-0.02$) leads to finite $T_C$ for $x>p$ or $x>c$, respectively. When $p$ is larger than half-filling of the correlated band $T_C$ decreases. A relative compensation $p=xc$ (green line, $p=x/2$) prefers ferromagnetism at every doping as long as $J$ is large enough (cf. Fig. \ref{figTcn}.) Symbols are from the experiment\cite{wang}. Parameters: $S=\frac{5}{2}$, $W=4$eV, $J=-1$eV, $T^{\text{NM}}_0=-0.5$eV}
\end{figure}
The first is a constant compensation $c$ resulting in a hole density $p=x-c$. For $x<c$ there is of course no ferromagnetism possible since the ordering mechanism is carrier mediated. With increasing doping the system has the optimal $p\approx x/3$ where the maximum $T_C$ is reached (cf. Fig. \ref{figTcp}). After that further impurities decrease the Curie temperature again. The second type of compensation shall result in a constant $p$. Also in this case we need a finite doping $x$ to get a positive $T_C$ since $x$ has to be larger than $p$. But with increasing $x$ the Curie temperature changes only slightly. Finally there could be a constant \emph{relative} compensation $p=cx$. We have chosen $c=\halb$ in this work. Thus the hole density is for all $x$ near its optimal value and there is a large finite $T_C$. Since there is the same effective filling $\gamma=p/x=c$ and the system is in the strong coupling limit, the dominant parameter is the effective bandwidth $W^{\text{eff}}\approx\sqrt{x}W$. The Curie temperature scales therewith almost only with the effectice bandwidth as it is usual in the strong coupling regime (cf. maxima of $T_C$ in Fig. \ref{figTcnvarx}). This results in the curvature $T_C\sim \sqrt x$.\\
Of course these compensation mechanisms are idealized and in reality there is probably a combination of these three basic types. But it can be seen that the hole compensation is indeed a very important effect for the DMS.

\section{Summary}

We have performed a model calculation to describe carrier mediated magnetic ordering of local moments. To do that we used a Kondo lattice model with ferro- or antiferromagnetic coupling $J$ between the carriers and the local moments. The electronic subsystem is treated in an equation of motion method to calculate the Green's function and the internal energy. From the internal energy we get the free energy which can be minimized according to the magnetizion of the local moments. This yields the actual magnetization of the system. To include disorder and dilution of magnetic impurities we have extended this method by a coherent potential approximation and a proper alloy analogy.\\
With this method we investigated the influence of several model parameters on the Curie temperature. We found that the optimal values are around quarter-filling of the conduction band in the strong coupling $|J|S>W$ regime. Additionally the free bandwidth $W$ should be as large as possible as long as the strong coupling limit is fullfilled. If $W$ is too large ($W>|J|S$) $T_C$ decreases. The effect of the bandwidth plays also a leading role in diluted systems. Since the effective bandwidth of the correlated band, which arises from the interaction of carriers and (diluted) local moments,  is proportional to the square root of the doping, $W^{\text{eff}}\approx\sqrt{x}W$, increasing $x$ enhances the Curie temperature. At half-filling of the correlated band no ferromagnetic ordering exist, which shows the importance of hole compensation in diluted magnetic semiconductors. It is also necessary for high $T_C$s that the correlated band is separated from the non-correlated band and that the chemical potential is within the correlated band. Finally we have compared the calculated values of $T_C$ with experimental ones of Ga$_{1-x}$Mn$_x$As which are in good agreement to each other.\\
Since we calculate the free energy it is possible to determine other quantities like the entropy or heat capacity of a system. Also free energies of different (magnetic) phases could be compared to see which phase is actually present at finite temperatures. We will leave this for future work.

\bibliographystyle{pss}
\bibliography{freeDMSv4}

\end{document}